\documentclass[preprint,showkeys,aps,prb]{revtex4}
\usepackage[dvips]{graphicx}
\begin{document}

\title{
What {\it is} Liquid ? [in two dimensions]
}

\author{
Karl Patrick Travis                              \\
Department of Materials Science \& Engineering   \\
The University of Sheffield                      \\
Sheffield S1 3JD, United Kingdom, and            \\
William Graham Hoover with Carol Griswold Hoover \\
Ruby Valley Research Institute                   \\
Highway Contract 60, Box 601                     \\
Ruby Valley, Nevada 89833, and                   \\
Amanda Bailey Hass                               \\
Department of Applied Maths                      \\
University of Leeds                              \\
Leeds LS2 9JT, United Kingdom                    \\
}

\date{\today}


\vspace{0.1cm}

\begin{abstract}
We consider the practicalities of defining, simulating, and characterizing ``Liquids''
from a pedagogical standpoint based on atomistic computer simulations. For simplicity and
clarity we study two-dimensional systems throughout. In addition to the infinite-ranged
Lennard-Jones 12/6 potential we consider two shorter-ranged families of pair potentials.
At zero pressure one of them includes just nearest neighbors. The other longer-ranged
family includes twelve additional neighbors. We find that these further neighbors can help
stabilize the liquid phase.

What about liquids? To implement Wikipedia's definition of liquids as conforming to their
container we begin by formulating and imposing smooth-container boundary conditions. To
encourage conformation further we add a vertical gravitational field. Gravity helps stabilize
the relatively vague liquid-gas interface. Gravity reduces the messiness associated with the
curiously-named ``spinodal'' (tensile) portion of the phase diagram. Our simulations are
mainly isothermal.  We control the kinetic temperature with Nos\'e-Hoover thermostating,
extracting or injecting heat so as to impose a mean kinetic temperature over time. Our
simulations stabilizing density gradients and the temperature provide critical-point
estimates fully consistent with previous efforts from free energy and Gibbs ensemble
simulations.  This agreement validates our approach.

\end{abstract}

\maketitle

\section{What is Liquid [ in two dimensions ] ? }

This work had its origin in the recent death of our colleague Douglas Henderson\cite{b1}.  Bill's
friendship with Doug dated back to the 1960s, their early years as scientists, working at the
Lawrence Livermore Radiation Laboratory (Bill) and IBM's Almaden Research Centre in San
Jos\'e (Doug). Bill and Carol visited Doug and RoseMarie's homes south of San Francisco and,
after the Loma Prieta earthquake of 17 October 1989, in Sandy Utah. These visits became more
frequent following the Hoovers' move to Ruby Valley Nevada in 2005.

All four of us authors have carried out research work devoted to a longstanding challenge of
equilibrium statistical mechanics, a better understanding of liquid state structure. The Mayers'
virial series for gases and the Einstein and Debye models for ordered solids provide a relatively
accurate understanding of matter's simplest pair of phases. Liquids remain more mysterious. The
question asked by Doug and John Barker in 1976\cite{b2} was a good one and remains so today. Bill
adopted this same title as the basis for two publications, one in 1998\cite{b3}, the second in
2014\cite{b4}, the latter as part of the celebration of Doug's 80th birthday.

\section{van der Waals' 1873 model for gases and liquids}

Though atomistic liquid structure remains mysterious, van der Waals provided us with his Nobel
Prize winning macroscopic ``equation of state''. This thermodynamic model describes both gases and liquids
as well as the ``critical'' condition at which the two become indistinguishable. van der Waals
chose two material properties to describe the strengths of the attractive and
repulsive contributions to the pressure and energy of fluids, both gaseous and liquid. For
simplicity we adopt ``reduced units'' here, setting van der Waals' material properties $a$,
characterizing attraction, and $b$, characterizing repulsion, both equal to unity. 
Throughout this work we use reduced units, setting particle masses and Boltzmann's constant
both equal to unity in addition to the potential parameters and van der Waals' $a$ and $b$.
In two space dimensions, with kinetic energy $K = \sum (p^2_x+p^2_y)/2 = NT$,  van der Waals'
mechanical and thermal equations of state are:
$$
(P + \rho^2)(1-\rho) = \rho T \ [ \ {\rm Mechanical} \ ] \ ; \
 e = T - \rho \ [ \ {\rm Thermal} \ ] \ .
$$
$P$, $\rho$, $T$, and $e$ --- pressure, density, temperature, and internal energy --- are the
macroscopic thermodynamic variables linked together by van der Waals in his 1873 dissertation
on {\it The Continuity of the Gas and Liquid States}.  For consistency with thermodynamics the
resulting mechanical and thermal properties of these fluids are correlated by the
second-derivative ``Maxwell relation'' that follows from the mixed partial derivatives of $[A/T]$
with respect to volume and temperature, $(\partial^2 [A/T]/\partial V \partial T) =
(\partial^2 [A/T]/\partial T \partial V)$
where $A$ is Helmholtz' free energy, $E - TS$ and $S$ is entropy:
$$
(\partial [P/T]/\partial T)_v = (\partial [e/T^2]/\partial v)_T = (\rho/T)^2 \ {\rm for \ van \ der \ Waals} \ .
$$
 
According to van der Waals' model and likewise in accord with nature, the gas and liquid phases can
only be distinguished at temperatures below a critical isotherm, on which the unstable minima and
maxima of van der Waals' pressure equation, a cubic in the density, coalesce. The ``critical
point'' on this isotherm is the only location in the pressure-density plane where the isothermal
slope and curvature simultaneously vanish:
$$
(\partial P/\partial \rho)_T = 0 \ {\rm and}  \ (\partial^2 P/\partial \rho^2)_T = 0 
 \longrightarrow \{ \ P_c=1/27, \ \rho_c = 1/3, \ T_c = 8/27 \ \} \ . 
$$
At this critical point the two fluid states, gas and liquid, become indistinguishable.
They also become hard to investigate as the vanishing first derivative implies infinite
compressibility, $-d\ln V/dP$, and zero sound speed, as $c = \sqrt{(\partial P/\partial \rho)_S}
\propto \sqrt{(\partial P/\partial \rho)_T}$, where $S$ is entropy. This singular behavior of the
pressure derivatives is reflected in the macroscopic nature of critical density fluctuations
big enough to see.  The fluctuations are observed visually as a milky ``critical opalescence''.

For any temperature less than the critical temperature value $T_c = (8/27)$ van der Waals'
model gives two values of the density which can coexist at mechanical and thermal
equilibrium. In addition to these ``binodal'' points there are two other density values
on every subcritical isotherm and on some adiabats between which the van der Waals equation of
state is mechanically unstable. These pairs of points form the high-density and low-density
boundaries of unstable isothermal and adiabatic ``spinodal regions'', within which at
least one of the two van der Waals' compressibilities is negative. Straightforward algebra
shows that van der Waals' adiabatic spinodal line [where $(\partial P/\partial \rho)_S$
vanishes] has the same form as the isothermal line, but at half the temperature :
$$
T_{\rm isothermal} = 2\rho(1-\rho)^2 = 2T_{\rm adiabatic} \ .
$$
 
The van der Waals model's ``binodal'' equilibrium coexistence curves and the ``spinodal''
curves of divergent compressibility are characteristic of many real macroscopic fluids and
common microscopic fluid models which include both attractive and repulsive pair forces. The
best known microscopic model is Lennard-Jones' 12/6 potential from the 1920s:
$$
\phi_{LJ}(r) = (1/r)^{12} - 2(1/r)^6 \longrightarrow
\phi^\prime(1) = 0 \ ; \ \phi(1) = - 1 \ .
$$
The van der Waals and Lennard-Jones phase diagrams are compared in {\bf Figure 1}. 
Although van der Waals' equation has no solid phase, a more sophisticated state-equation model,
based on the known hard-disk equation of state, for disks of diameter $\sigma$ and number
density $\rho = (N/V)$:
$$
(PV/NkT) = (PV/NkT)_{\rm disks} - \rho \ \ [ \ {\rm with} \ B_2 = (\pi/2) \ {\rm and \ diameter} \
\sigma \equiv 1 \ {\rm for \ disks} \ ] \ ,
$$
provides a three-phase equation of state analogous to van der Waals' two-state solution. The
critical parameters depend upon the reduced units chosen
for the hard-disk model. With disk diameter $\sigma$ and Boltzmann's constant set equal to
unity this model gives $(P_c, \ \rho_c, \ T_c) = (0.019, \ 0.269, \ 0.216)$ with a dimensionless
compressibility factor $(PV/NkT) = 0.326$, quite close to the value of (1/3) obtained by using a
maximally truncated three-term virial expansion, $(PV/NkT) = 1 + B_2\rho + B_3\rho^2$.

\begin{figure}
\includegraphics[width=3. in,angle=-90.]{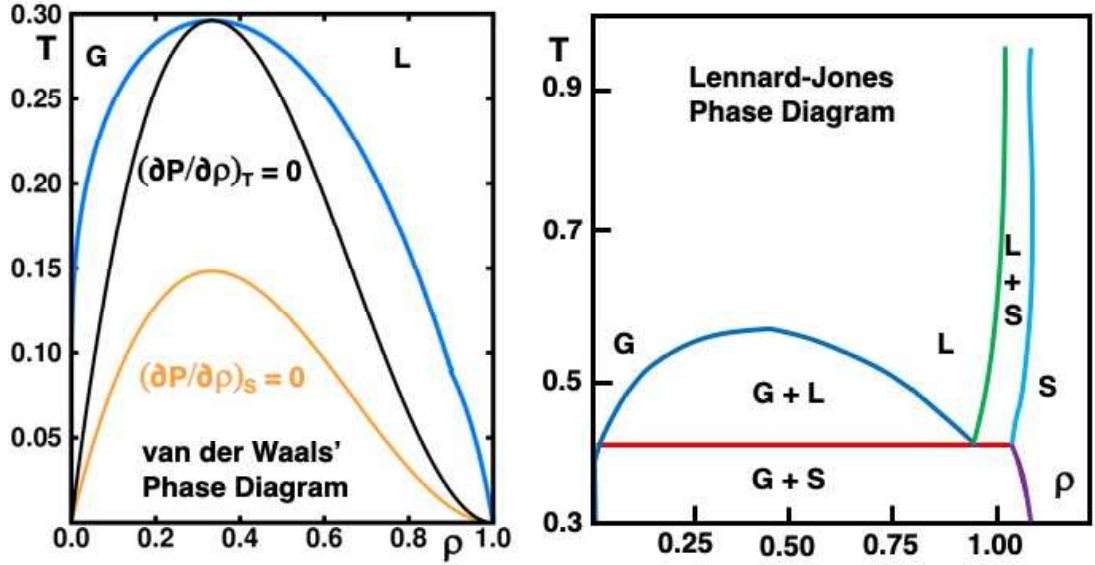}
\caption{
At the left is van der Waals' phase diagram and at the right the two-dimensional Lennard-Jones
analog\cite{b5}. In the van der Waals case the isothermal and isentropic spinodal lines, where
the corresponding compressibility diverges, are shown. In the Lennard-Jones case, with its solid
phase, there is a triple point near $T=0.4$ below which the gas and solid coexist. Between the
triple-point temperature and the critical temperature (roughly 0.56) lower-density gas and
higher-density liquid can coexist.
}
\end{figure}

It is evident from {\bf Figure 1} that pair potential models can provide a semi-quantitative
understanding of the coexistence and coalescence of the less-dense gas and more-dense
liquid phases of simple fluids, where both phases are formed from the same ingredients.

Understanding the details of the microscopic structure leading to this macroscopic behavior is
an excellent illustration of the problem areas all four of us authors have enjoyed exploring.
Before entering into the details of our own work let us consider the progressive steps leading
from van der Waals equation in the late 1800s up to Barker and Henderson's review a century later.

\section{Theories and Models of the Liquid State}

In the late nineteenth century critical-point experiments were carried out by heating a known
quantity of liquid in a sealed tube with an obvious ``meniscus''. That word, ``meniscus'', comes
from the Greek for ``curved moon''.  It is because the two phases interact with their container's
surface differently that the meniscus separating them is curved.   Near the ``critical point'',
where gas and liquid become indistinguishable, dramatic density fluctuations
broaden and destroy the meniscus separating the two coexisting phases.

In 1882 Hannay declared ``The formation of a meniscus is the only test of the liquid state''\cite{b6}.
That meniscus definition is superior to Wikipedia's notion, ``Liquid is a nearly incompressible
fluid that conforms to the shape of its container''. Hannay was right. A liquid-gas interface,
or meniscus, with the liquid the denser of the two fluid phases, is necessary to distinguish
the one phase from the other. Wikipedia's liquid definition would include dense fluids of hard
disks or spheres.  But neither of those hard-particle systems has the attractive forces
necessary to stabilize a liquid phase.

By 1900, with the advent of Boltzmann and Gibbs' statistical mechanics, atomistic models became
important. Kinetic theory and lattice dynamics offered useful descriptions of gases and solids.
Good structural models for liquids were absent. This lack soon motivated the construction of
physical models of liquid structure. In 1930s London John Bernal simply added more and more ball-bearing
particles to ball-and-stick or conglomerate balls-in-paint structures. Bernal found that the
radial distributions of pairs of balls resembled those inferred from radiation experiments on
real liquids.  At about this same time Joel Hildebrand, in Berkeley, immersed more than 100 gelatin balls in a fish
tank, likewise finding that the distribution of the balls' separations resembled the
distributions of interparticle distances in liquid argon, scaled up by eight-or-so orders of
magnitude.  Bernal and Hildebrand were looking for bulk liquid structure, not the interfacial menisci
stressed by Hannay.

John Barker devoted most of his working life to the understanding of liquids\cite{b7}, publishing his only
book, {\it Lattice Theories of the Liquid State} in 1963, just as it was becoming clear that lattices
were not a proper starting point for ``understanding''.  Before he and Doug Henderson had discovered
and implemented  perturbation theory Barker had attempted to improve his understanding of liquids by
extending ``lattice theories'', like the Ising Model. That model, with its hole-particle symmetry,
seems very distant to real liquids.  Barker invented ``tunnel models'', taking advantage of the
mathematical simplicity of one-dimensional chains of particles, coupled with a symmetry-breaking
description of the tunnel locations. In tunnel models for simple atoms one-third of the degrees of
freedom are longitudinal and two-thirds are transverse to the tunnels.  Barker's collaboration with Doug
in the 1960s, based on a perturbation theory of the Helmholtz free energy, was soon to provide a
surprisingly useful predictive theory.  The theory provided all the liquid thermodynamic properties based
on known properties of hard spheres. This hard-sphere-based theory's success seems a bit puzzling because
the underlying model is itself incapable of providing the two-fluid meniscus characteristic of real
 liquid-gas coexistence.

\section{Progress in Understanding from Computer Simulations}

Soon after World War II, in the 1950s, the advent of computers opened up completely new research
opportunities. Alder, Jacobson, Wainwright, and Wood developed Monte Carlo and molecular dynamics
simulation algorithms modeling equilibrium distributions of dozens or hundreds of hard particles in two
and three space dimensions. They discovered and characterized the hard-disk and hard-sphere fluid-solid
transitions\cite{b8,b9}. These melting-freezing transitions occur when the solid phases are expanded about ten
percent in $(x,y)$ or $(x,y,z)$, corresponding to melting densities, relative to close-packed, of about
(4/5) (for disks) and (3/4) (for spheres).

In 1958 Jerry Percus and George Yevick formulated an integral equation for the pair distribution
function\cite{b10}. Mike Wertheim solved the equation analytically for hard spheres five years
later\cite{b11}. A numerical solution of the hard-disk analog appeared  half a century later, in 2008.
The analytic work for spheres gave an excellent approximation to the hard-sphere distributions from
Monte Carlo and molecular dynamics simulations. These developments led to the successful refinements
of perturbation theory reviewed by Barker and Henderson in 1976. Their approach was paralleled by
several other dedicated scientists, among them Farid Abraham, Hans Andersen, Frank Canfield, David
Chandler, Ali Mansoori, Jay Rasaiah, George Stell, and John Weeks.

\section{Barker and Henderson's Description of ``Liquids''}

After a decade working together Barker and Henderson addressed our title question from the standpoint
of perturbation theory, in 1976. Rather than constructing physical many-body models they
adopted the results of hard-particle computer simulations to develop and evaluate a perturbation
theory based on the Percus-Yevick hard-sphere distribution function. They treated attractive forces
as a perturbation added to a reference repulsive potential. The resulting free energy calculations
related the thermodynamics of homogeneous liquids to hard-sphere fluid-phase properties. Helmholtz'
and Gibbs' free energies can alternatively be found by integrating equation of state data taken from
Monte Carlo or molecular dynamics simulations. With today's computers brute-force simulation is the
more practical and much-simpler approach.

Barker and Henderson summarized the state of the art of the 1970s perturbation work in their review. In
its simplest form liquid perturbation theory is based on optimizing a reference-system's hard core size
by minimizing Helmholtz' free energy at fixed values of the density and temperature. The success of this
theory is due to the fact that thermodynamics requires no treatment of mixed-phase systems. Consequently
perturbation theory can be based on reference hard-particle systems which lack a liquid phase and its
corresponding meniscus. Bill Wood, at Los Alamos, pointed out that the hard-particle systems' fluid-solid
surface tension is negative.  Thus drops of hard disks and spheres don't form. Unlike models with
attractive forces hard particles don't form clusters.

\section{Conceptual Difficulties : Liquids' ``Spinodal Region''}

There is a tremendous literature on the ``spinodal'' region of the phase diagram\cite{b12,b13,b14}. For
van der Waals' equation this is usually taken to be the mechanically-unstable region with negative
isothermal compressibility. In principle a negative compressibility, either isothermal or adiabatic,
generates exponential growth of density fluctuations and so is to be prohibited in realistic fluid models.
Thus the borders of a spinodal region for real fluids, if there were one, would be hard to access and
describe. Wedekind {\it et alii} described access to the spinodal regions well\cite{b13}:\\
\begin{small}
\begin{center}
``It would be very complicated, if not practically impossible, to reach the spinodal.''\\
\end{center}
\end{small}

The internet reveals that ``spinodal'' originated as a synonym for ``cusp''. This explanation seems
curiously incomplete (and mercifully absent from most textbooks) as no cusp is apparent in realistic
phase diagrams like those of van der Waals or the Lennard-Jones potential.  See again {\bf Figure 1}. We
consider a region with negative compressibility strange, artificial, and ``unstable''.  Shamsundar and
Lienhard explicitly object\cite{b12} to the term ``unstable'', citing the reality of nearby states of
superheated liquids and supercooled gases, likewise nonequilibrium states not appearing in a 
conventional single-valued phase diagram. Any {\it fluid} under tension and subject to mechanical noise
cannot persist unchanged for long. Solids do characteristically exhibit  tensile strength but still can
suffer shear instabilities. We will illustrate such models later in this work. There is as yet no accepted
standard for bulk ``liquid'' structure, though not for lack of trying.

In the context of the usual single-component thermodynamics for a simple material like argon
the spinodal region of the phase diagram corresponds, at least conceptually, if not in the laboratory, to a mix of a
denser liquid and a less-dense gas. Because such a system isn't homogeneous it is clear that a simple phase
diagram or a model like van der Waals' is an incomplete description. Simulations in this region lead to
highly-complex evolving structures of transient rotating clusters or clumps of particles. The properties
of nonequilibrium clusters are complex, involving surface tension and rotational contributions to the
energy, making the characterization of ``pressure'' somewhat uncertain.

In our own effort to clarify the ambiguities of ``spinodal states'' for two-dimensional fluids we thought
it prudent to consider three different initial conditions, all of them equally plausible {\it a priori}. With
Lennard-Jones forces both
the square lattice, sufficiently expanded to reach a density of 0.4, or a triangular lattice expanded to that
same density, have energies exceeding that at the critical point and so cannot serve as models for a spinodal
state using conservative mechanics. A simple way out of this energy problem is to consider {\it isothermal}
molecular dynamics, starting and finishing at an imposed kinetic temperature less than critical and greater
than that at the triple point. Such a choice lies somewhere in a nonequilibrium liquid range. In two space
dimensions, the Lennard-Jones critical and triple-point temperatures are on the order of 0.56 and 0.4 according
to Barker, Henderson, and Abraham\cite{b5}. We have chosen the temperature 0.5 as our standard initial (and final,
time-averaged) condition for our exploratory molecular dynamics simulations. See {\bf Figure 2} for a sample
evolution from an unstable square lattice. An expanded triangular lattice provides a similar history. A third 
possibility, illustrated in {\bf Figure 3} is to divide up the system into cells with the structure in each
cell chosen randomly. A special case of this choice takes a regular stress-free lattice with the number of
randomly-located vacancies chosen to satisfy the desired density, 0.4 in our case, close to the
Barker-Henderson-Abraham estimate of the critical density in {\bf Figure 1}.

\begin{figure}
\includegraphics[width=3. in,angle=-90.]{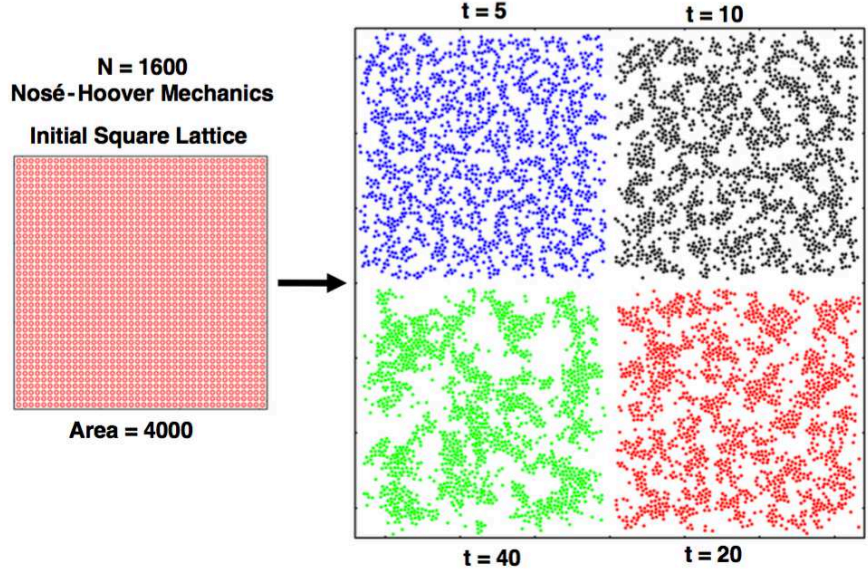}
\caption{
Four snapshots in a ``spinodal'' evolution. The dynamics is Nos\'e-Hoover isothermal at the subcritical
temperature $T = 0.5$. The initial condition, at the left, is a perfect square lattice of area 4000 containing 
1600 Lennard-Jones particles. Fourth-order Runge-Kutta integration with $dt = 0.005$ to time 40. The
instability of the lattice gives rise to coarsening, soon forming a percolating cluster spanning the
volume\cite{b14}. Boundary potentials quartic in $dx$ and $dy$ repel any particles with $| \ x \  |$ or
$| \ y \  |$ exceeding $\sqrt{1000} = 31.623$.
}
\end{figure}

\begin{figure}
\includegraphics[width=3. in,angle=-90.]{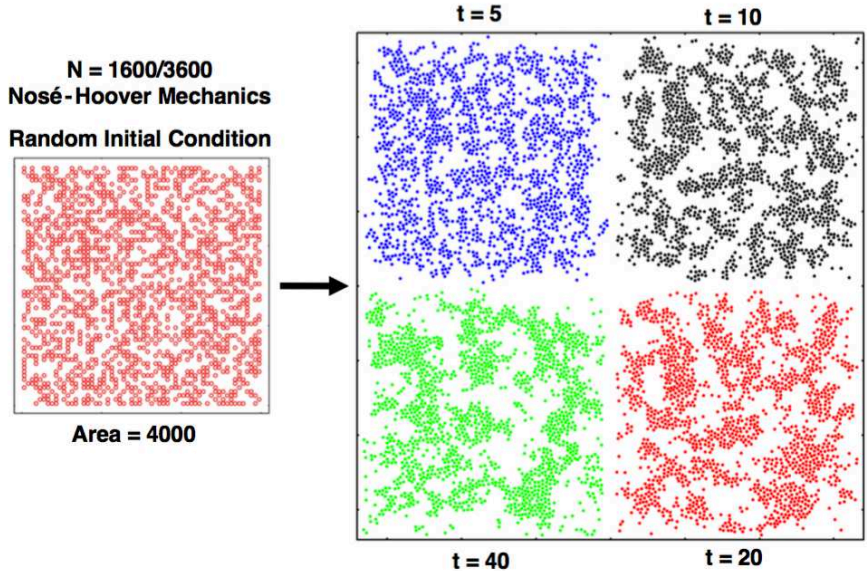}
\caption{
Four snapshots in a ``spinodal'' evolution. The dynamics is Nos\'e-Hoover isothermal at the subcritical
temperature $T = 0.5$. The initial condition was a perfect square lattice of 3600 sites in an area 4000
with 1600 of the sites, randomly chosen, containing Lennard-Jones particles. Boundary potentials quartic
in $dx$ and $dy$ repel any particles with $| \ x \ |$ or $| \ y \ |$ exceeding $\sqrt{(1000)} = 31.623$.
}
\end{figure}

The two sample evolutions shown here are typical of the spinodal region. The equilibrium phase diagram disallows states
under tension. There is an initial exponential growth of density fluctuations, followed by a slower coarsening
of clusters to form a percolating cluster spanning the entire volume\cite{b14}. The details of the first exponentially
unstable phase depend upon the initial conditions. The details of the second phase, with the nonequilibrium
equilibration of growing clusters, are relatively easy to see but hard to predict\cite{b15}, suggesting the exploration of
alternative methods for characterizing liquids.

Gravity can take us in the direction of Hannay's meniscus. The
evolution of an initial state toward the formation of a meniscus can be visualized by adding a small
gravitational field to the dynamics.  With a field the liquid state lies below the vapour with
which it equilibrates. The {\bf Figure 1} phase diagram for Lennard-Jones' potential indicates that a
liquid about six times denser than its vapour should be stable at a temperature of 0.5, well below the
critical temperature of 0.56 and above the triple point temperature of 0.4. This suggests a feedback
dynamics similar to the ``Gibbs ensemble'' algorithm, with particles transferred from interacting
simulations with a common pressure and temperature. Feedback within a single simulation provides a less
singular evolution. Let us turn to dynamics in the presence of an external gravitational field.

\section{Implementing A ``Liquid'' Vision with Molecular Dynamics}

In the present work we consider the need, and fill it, for a ``meniscus'' separating a ``gas'' from a
``liquid'' fluid. We stabilize and investigate the interfaces defining phase boundaries. To do
this we first of all model the idea of a physical ``container'', to which all the particles in our
simulations must conform. To simulate the dynamics of such a manybody system we enclose it within a
special fixed boundary, a smooth nearly-rigid container modeled with a quartic repulsive surface
potential.

In keeping with the expected accuracy of a fourth-order Runge-Kutta integration of the motion equations
we adopted one-sided quartic potentials to contain our simulations.  We begin with both rectangular  and
circular ``containers'' for our molecular dynamics. The boundary potential energy in the
circular case is $(dr^4/4)$ where $dr$ is the depth of penetration beyond a circular boundary of radius
$r$. To reflect escaping particles in the rectangular case the boundary potentials are $(dx^4/4)$ and
$(dy^4/4)$ imposed on the four sides of a rectangular container. $dx$ and $dy$ are the penetrations
beyond the vertical and horizontal walls of a rectangular container. See {\bf Figures 4 and 5} for
typical equilibrium snapshots of 400 Lennard-Jones particles with the density and kinetic temperature at
a fluid state point well above the gas-liquid coexistence curve, $\rho = 0.4; \ T = 1 > T_C
\simeq 0.56$.  The wild density fluctuations seen in these two equilibrium snapshots rightly suggest that
time-averaging is needed to aid the analysis of the gas-liquid meniscus structure. 
 
\begin{figure}
\includegraphics[width=3. in,angle=-90.]{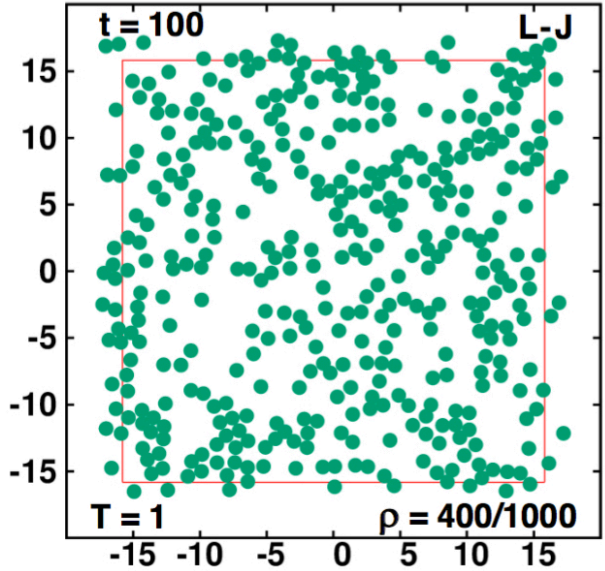}
\caption{
An equilibrium snapshot of 400 Lennard-Jones particles at $T=1$ confined by square boundary potentials at
$\pm \sqrt{250} = 15.811$. The density is 0.4. The simulation time $t = 100$ is adequate for equilibration.
Here, and mostly throughout, we use fourth-order Runge-Kutta integration with a timestep $dt = 0.005$.
}
\end{figure}

\begin{figure}
\includegraphics[width=3. in,angle=-90.]{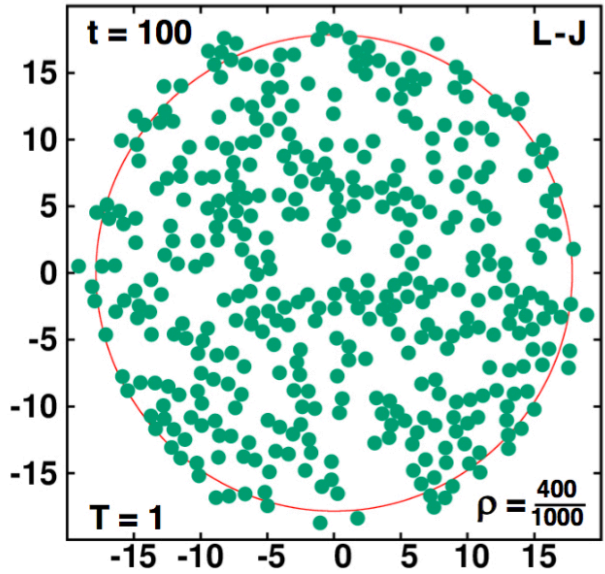}
\caption{
An equilibrium snapshot of 400 Lennard-Jones particles at $T=1$ confined by a circular quartic boundary
potential at $r = \sqrt{1000/\pi} = 17.841$. The snapshot was taken after 20,000 timesteps with $dt = 0.005$.
The density is 0.4. The simulation time $t = 100$ is adequate for equilibration.
}
\end{figure}

The figures document that typical penetrations, beyond the quartic boundaries, are about one
particle diameter, consistent with an energy-based estimate :
$$
(dx^4/4) \simeq (dy^4/4) \simeq (dr^4/4) \simeq T = 1 \rightarrow dr \simeq 1.414 \ .
$$
With these straightforward model boundary potentials providing a conforming container we next seek out a
means for emphasizing and localizing the meniscus characteristic of the liquid state.

\section{Stabilizing the Meniscus with Gravity}

Barker  and Henderson were satisfied with a formal semiquantitative perturbation theory based
on a reference hard-core potential. We prefer a more physical approach, based on observations
of phase equilibria. For us, a stable interface separating a liquid from its less-dense gas is the
necessary and defining aspect of liquid behavior. From the observational standpoint to
be sure one is viewing a liquid (as opposed to a gas or hard-particle fluid) requires observing
the interface separating the two varieties of simple fluids, the liquid and the gas. This is easy
to do by simulating a gas bubble surrounded by liquid or a liquid drop in a dilute gas; but such
clearcut obervations become blurred nearer the ``critical point'' where fluctuations are macroscopic.
There the manybody dynamics is dominated by percolating clusters of macroscopic size.

To encourage our particles' conformation to their container with a visible meniscus we include a
second innovation, a constant vertical acceleration, $-g$ for each of our computational particles.
This constant downward force is added to the pairwise forces from other particles and to the boundary
forces defining our containers :
$$
F = F_{pair} + F_{penetration} + F_{gravity} \ .
$$
We harbor the optimistic assumption that such a combination of particle plus boundary plus gravitational
forces will accommodate not only a gas-liquid interface but also the liquid-solid one.  In 1977 Ladd and
Woodcock demonstrated that sufficiently close to the triple point it is possible to see both of these
liquid interfaces simultaneously\cite{b16}.  At such a ``triple point'' there are no thermodynamic ``degrees of
freedom''. All three phases coexist at the same pressure and temperature. By adding gravity we provide our
fluid systems with a pressure gradient satisfying the continuum force balance, $(dP/dy) = -\rho g$ in the 
stationary state.  The pressure gradient $(dP/dy)$ forces the fluid to conform its shape to its container,
and, over a wide range of pressures, serves to localize and illustrate the gas-liquid and liquid-solid
interfaces.

The phenomena of yield stress (for the solid) and surface tension (for the liquid) could prevent shape
conformation unless these properties can be overcome by gravitational or rotational forces. We have chosen
gravity as the simpler of these two choices.  Finally, in order to prescribe the overall temperature of our
two-phase or three-phase systems we apply Nos\'e-Hoover isothermal dynamics with a target temperature
$T_{NH}$ :
$$
T_{NH} = \langle \ (K/N) \ \rangle = \langle \ (p_x^2/2) + (p_y^2/2) \ \rangle \ .
$$

\section{Imposing Kinetic Temperature on Molecular Dynamics}

The fundamental conceptual  basis of our present work is conservative Hamiltonian molecular dynamics.
We include the forces and potential energies from a constant gravitational field as well as those
describing special containerized boundary conditions. For flexible control of the simulations, and to
accelerate convergence, we generalize the underlying mechanics to include Nos\'e-Hoover control of
temperature.  Let us next outline the Nos\'e-Hoover control mechanism.

Molecular Dynamics with specified kinetic temperatures has made steady-state nonequilibrium simulations
a standard method for simulating steady nonequilibrium flows of mass, momentum, and energy. In 1984 Nos\'e
introduced his novel time-reversible Hamiltonian dynamics. He treated the kinetic temperature as an
independent variable imposed on the dynamics. This is accomplished by augmenting the  manybody motion
equations with a time-reversible friction coefficient $\zeta$. Hoover provided a simplified formulation
of Nos\'e's approach which has been widely adopted.  We use it here:
$$
\{ \ \dot x = p_x \ ; \ \dot p_x = F_x - \zeta p_x \ ;
   \ \dot y = p_y \ ; \ \dot p_y = F_y - \zeta p_y \ \}      
$$
$$
\ \dot \zeta (t) =
(1/N\tau^2)[ \ K(t) - \langle \ K \ \rangle \ ] \ [ \ {\rm Nos\acute{e}-Hoover \ Dynamics} \ ] \ .
$$
Here $\langle \ K \ \rangle$ is the constant target value of the kinetic energy, imposed by $\zeta$.
In the two-dimensional systems we consider the kinetic temperature is $T=(1/N)\sum(p_x^2+p_y^2)/2 =
(K/N)$, and the relaxation time $\tau$ can be chosen as a typical collision time. For the current
simulations we have chosen $\tau = 1$.

If it is desirable to accelerate convergence it is quite practical to begin with a higher imposed
temperature and/or a higher gravitational field.  With Runge-Kutta integration it is perfectly feasible
to specify analytic time dependences for these target temperatures and fields, $T(t)$ and  $g(t)$,
within the equations of motion.

For simplicity and clarity we restrict our investigations to two-dimensional systems. We look directly
at coexisting phases, so as to avoid the need for free energy calculations. Corresponding implementations
for three-dimensional systems are straightforward.  This will be clear as we discuss the necessary
diagnostics for analyzing the results of our computer simulations.

\section{Isothermal Lennard-Jones Fluids with Gravity}

\begin{figure}
\includegraphics[width=3. in,angle=-90.]{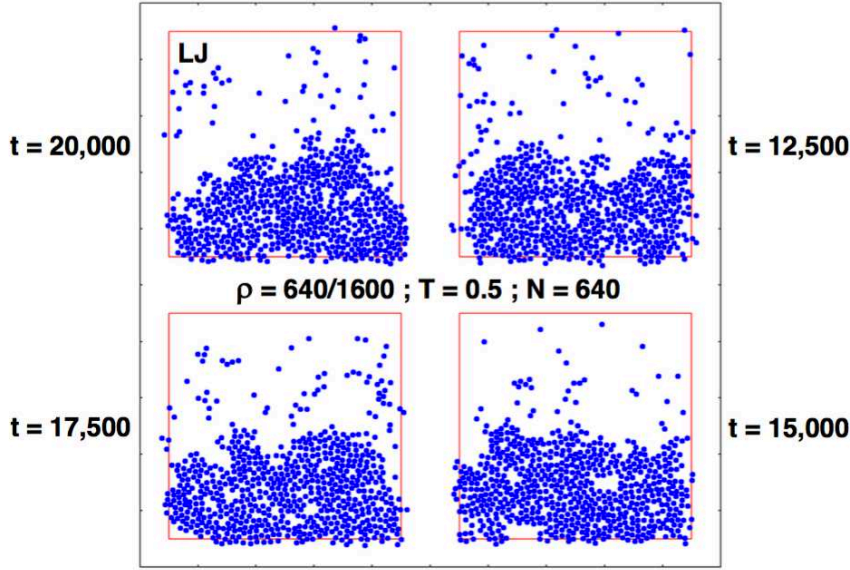}
\caption{
Four Nos\'e-Hoover Lennard-Jones snapshots at times 12,500, 15,000, 17,500, and 20,000. Gravitational field
0.01 with 640 particles confined by a square boundary, quartic potentials at $| \ x \ |$ = 20 and $| \ y \ |$ = 20.
Timestep 0.005 and $T = 0.5$. Fourth-order Runge-Kutta integration.
}
\end{figure}

\begin{figure}
\includegraphics[width=3. in,angle=-90.]{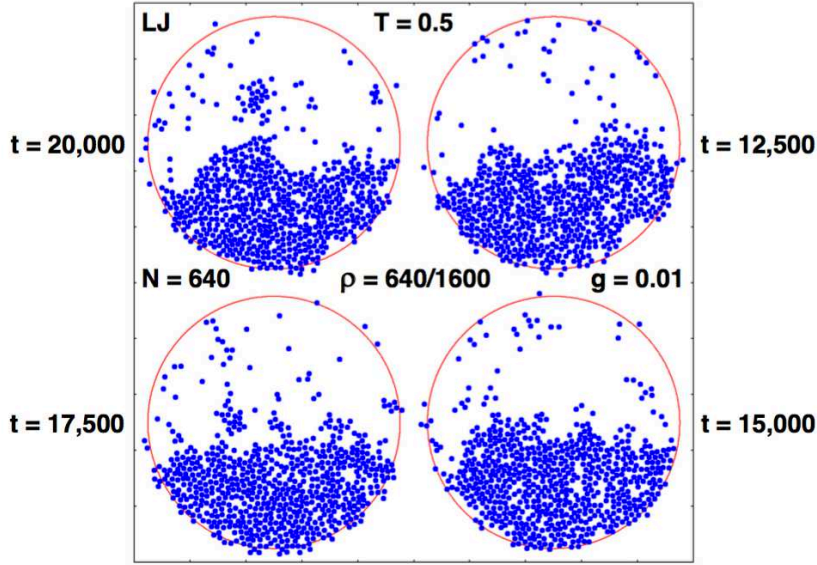}
\caption{
Four Nos\'e-Hoover Lennard-Jones snapshots at times 12,500, 15,000, 17,500, and 20,000. Gravitational field 0.01
with 640 particles confined by a circular boundary at $r = \sqrt{x^2 + y^2} = \sqrt{1600/\pi} = 22.568$. Timestep
$dt = 0.005$ and $T = 0.5$.
}
\end{figure}

The simulations leading to {\bf Figures 4 and 5}, when time-averaged, correspond to equilibrium
homogeneous fluids due to the lack of any organizing field. {\bf Figures 6 and 7}, which include gravity,
illustrate very different situations incorporating menisci. They snapshot the evolving morphology of 640
Lennard-Jones particles in square and round containers of enclosed volume 1600, corresponding again to a
near-critical density $\rho = 0.4$. In both these highly inhomogeneous systems the gravitational field
strength is $g = 0.01$. The slightly subcritical kinetic temperature, $\langle \ (p_x^2+p_y^2)/2 \ \rangle$,
is 0.5, and the Nos\'e-Hoover relaxation time imposing it is unity. The underlying two-million-timestep
simulations including all $(640\times639/2)$ Lennard-Jones interactions take a half day on a desktop
computer. The two field-driven figures, with four sample snapshots from the last halves of the runs, both
show a denser phase $\rho \simeq 1$ below a lower-density gas phase $\rho \simeq 0.1$. We will see in
{\bf Figure 17} that time-averaging isothermal snapshots provides additional simplicity and considerable
clarity.

Koch, Desai, and Abraham's comprehensive spinodal work using the Lennard-Jones potential\cite{b14} suggests
a gas-liquid density ratio of roughly (1/6) at $T=0.5$.  In our earlier exploratory simulations the gravitational
field strength obeyed a feedback differential equation based on generating our desired factor of six in the
density difference, $\dot g = 0.01[ \ N^{gas}(t) - (N/7) \ ]$. $N^{gas}$ is simply the number of
particles with positive $y$ coordinates. Because the resulting fluctuating field strength was
close to 0.01 we adopted the simpler and smoother approach of using a constant gravitational field for
{\bf Figures 6 and 7}. With cartesian coordinates $\{ \ q = (x,y), \ p = (p_x,p_y) \ \}$ the $4N+1$
differential equations of motion are:
$$
\{ \ \dot q = p \ ; \ \dot p = F - \zeta p \ \} \ ;
 \ \dot \zeta = (K/N) - 0.5 \ ; \ K \equiv \sum (p_x^2 + p_y^2)/2 \ .
$$
The summed forces $F$ on each particle include pair forces, gravity, and the container forces.  We add on the
thermostat forces, $\{ \ -\zeta p \ \}$, assigned the task of imposing isothermal conditions throughout the
container. The particle interactions are Lennard-Jones without any cutoff, with quartic boundary potentials on
the sides of the square container and along the perimeter of the circular container.

Individual densities $\{ \ \rho_i \ \}$, at each particle or at any grid point $(x_g,y_g)$, can be defined, and
evaluated numerically, with the help of Leon Lucy's two-dimensional ``smooth-particle'' weight function\cite{b17}.
The weight function spreads the influence of each particle very smoothly (two continuous derivatives everywhere)
in space. For example the ``delta-function'' density of each two-dimensional particle and of its properties (such
as velocity, energy, and pressure tensor) are likewise distributed smoothly within a circle of radius $h$.  The
weight function is maximum at the particle's location and vanishes on and beyond its bounding circle.  The
corresponding density distribution in the differential neighborhood $rdrd\theta$ of a particle is
$$
\rho(r,\theta) = (5/\pi h^2)(1 - 6z^2 + 8z^3 - 3z^4) \ ; \ z \equiv r/h \ .
$$
A reasonable choice of the range $h$ of Lucy's weight function for most atomistic simulations is 2 or 3
particle diameters.  We have chosen 2 throughout the present work.

The normalization prefactor in one dimension, $(5/4h)$ for $-h < dx < +h$, is replaced by $(5/\pi h^2)$
for normalization within a circular area with $\pi r^2 < \pi h^2$ :
$$
\int_0^h 2\pi rdr(5/\pi h^2)(1 - 6z^2 + 8z^3 - 3z^4) \equiv 1 \ {\rm where \ again} \ z \equiv (r/h) \ .
$$
Lucy's smooth weight function is convenient for comparing the results of atomistic simulations to the
predictions of continuum mechanics, as we shall presently demonstrate, when seeking interfaces
identifying the liquid phase.

A simple one-dimensional example illustrates the usefulness and power of smooth-particle weighting. Consider
the one-dimensional lattice of points at the integers so that the coarse-grained ``density of points'' is
unity.  Using Lucy's smooth-particle weighting function normalized for one-dimensional distributions,
$$
w(|dx|<h) = (5/4h)(1 - 6z^2 + 8z^3 - 3z^4) \ {\rm where} \ z \equiv |dx|/h \ ,
$$
gives for the density at each integer point 1.0156 for a ``smoothing length'' $h=2$ and 1.0031 for smoothing
length $h=3$. The local density in a one-dimensional system at the grid point $x_g$ is the summed-up
contribution from nearby particles $\{ \ x_i \ \}$ within a distance $h$ of the grid point:
$$
\rho(x_g) \equiv \sum_i w(| \ x_i - x_g \ |) \ .
$$
Applying this same definition, in between the integers, at the various midpoints $\pm(1/2)$, $\pm (3/2)$, 
$\dots$, the smoothed densities are 0.9863 and 0.9973 respectively for smoothing lengths of 2 and 3.

Likewise, carrying out a one-dimensional average over $x$ for a few hundred horizontal strips the
$y$-dependent pressure and density, $\{ \ \langle \ P(y) \ \rangle \ \}, \ \{ \ \langle \ \rho(y) \ \rangle
\ \}$ can be computed with one-dimensional weights including all particles within a vertical separation $|dy|<h$
of the gridpoint in question, where $h$ is the range of the Lucy function. The spatial and temporal averaging
process involves three distinct steps: At each timestep [1] Compute individual particle properties such as $\rho_i$
and $P_i$ using two-dimensional smoothing; [2] Convert the particle data into spatial averages for a $y$ grid
using one-dimensional smoothing; [3] Combine the spatial values by averaging over as many as millions of timesteps.

{\bf Figures 8 and 9} show time-averaged pressure and density profiles for the two boundary conditions, square
and circular.  In both cases wildly fluctuating snapshot configurations, time-averaged over the
last half of a two-million timestep simulation, provide a smooth meniscus with a width of just a few particle
diameters.  The pressure and density, averaged over both time and $x$ indicate a horizontal isotherm rather
than a van der Waals' loop.

\begin{figure}
\includegraphics[width=3. in,angle=-90.]{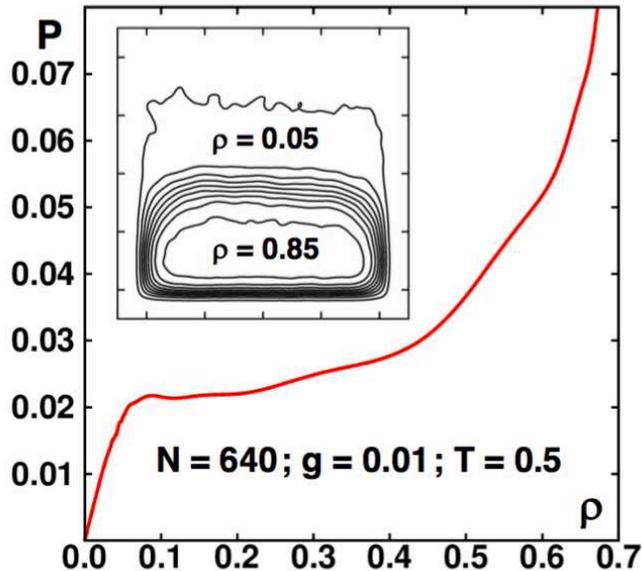}
\caption{
Time-averaged density contours from 0.05 to 0.85 for 640 Lennard-Jones particles at $T=0.5$ and overall
density 0.4. The quartic square boundaries are located at $\pm 20$. Space and time averages, over the
horizontal $x$ coordinate in space, and using the final two million timesteps in a four million timestep
run, in time, provide the $\langle \ P \ \rangle ( \ \langle \ \rho \ \rangle \ )$ profile giving the
structure of the meniscus perpendicular to that interface. 
}
\end{figure}

\begin{figure}
\includegraphics[width=3. in,angle = -90]{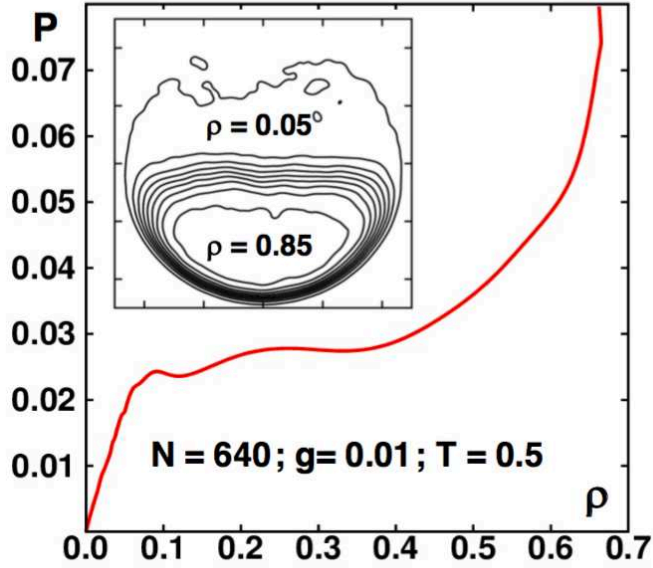}
\caption{
Time-averaged density contours from 0.05 to 0.85 for 640 Lennard-Jones particles at $T=0.5$ with overall
density 0.4. The circular boundary potential begins at $r=\sqrt{(1600/\pi)} = 22.568$. Averages over the
horizontal $x$ coordinate using two million timesteps provide the $\langle \ P \ \rangle (\langle \ \rho \ \rangle)$
profile giving the structure of the meniscus.
}
\end{figure}

\begin{figure}
\includegraphics[width=3. in,angle=-90.]{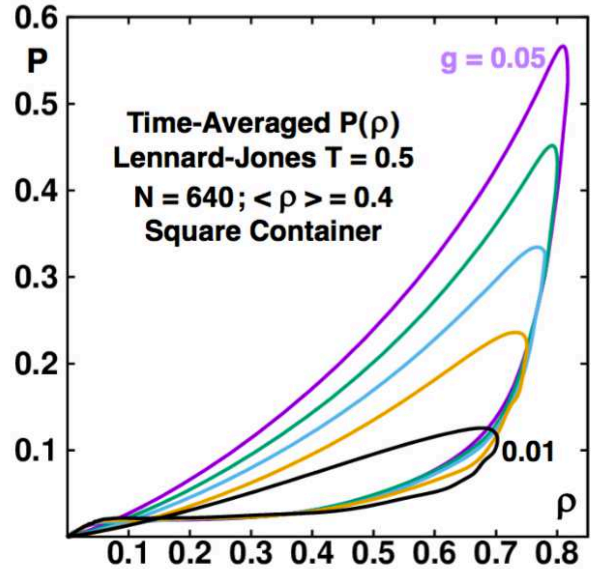}
\caption{
Time-averaged pressure as a function of time-averaged density using the last half of two million Runge-Kutta
timesteps for 640 Lennard-Jones particles in a $40 \times 40$ quartic box.  Five separate curves are shown
corresponding to field strengths 0.01, 0.02, 0.03, 0.04, and 0.05.  The lower portions of the five curves
correspond to the meniscus separating liquid from gas.  The good agreement indicates very little dependence
of the pressure-density correlation upon field strength. The simulation with $g = 0.01$ used the last half
of a four-million-timestep run.
}
\end{figure}

{\bf Figure 10} compares the structures of the meniscus, pressure versus density, for a series of five
values of the gravitational field from 0.01 to 0.05. The upper segment of each of the five traces
corresponds to the high-density high-pressure region near the bottom of the container, which supports
the entire weight of the 640 Lennard-Jones particles. The low-density low-pressure region near the
bottom of the plot (where the five traces agree) describes the meniscus atop most of the fluid. The
good agreement of all five indicates that the present introduction of gravity into critical-region
simulations provides accurate unambiguous estimates of the subcritical isotherms without the need for
free energies or a Maxwell construction. Contour plots, as in {\bf Figures 8 and 9} are probably the
best diagnostic tool for the meniscus as the density near the centre of the container can be assessed
and, when time-averaged, is guaranteed to obey the barometer formula, $dP(y)/dy = -\rho(y) g$.

\section{Two Finite-Range Polynomial Pair-Potential Families}

In our exploratory work here we have emphasized Lennard-Jones' 12/6 pair potential because its
thermodynamic properties are familiar and well investigated\cite{b18,b19}. Lennard-Jones' potential
is the most thoroughly studied of the ``realistic'' potentials. We have sought to learn more by
introducing two very different families of finite-ranged pair potentials. With a triangular lattice
the longer-ranged family
$\{ \ \phi^L_m(r<2) \ \}$, includes three shells of neighbors, 18 in all, at zero stress, while the
short-ranged potentials, $\{ \ \phi^S_m(r<\sqrt{2}) \ \}$ include only the six nearest neighbors:
$$
\phi^L_m = (2-r)^{2m} - 2(2-r)^m \ {\rm for} \ r < 2 \ ; \
\phi^S_m = (2-r^2)^{2m} -2(2-r^2)^m \ {\rm for} \ r < \sqrt{2} \ .
$$
See {\bf Figure 11} for six example plots of these relatively short-ranged potentials. For
all of them we will continue to adopt ``reduced units'' based on a well-depth of unity at the
particle-pair separation of unity. All the potentials have smooth minima of $-1$ at $r = 1$. In the
Lennard-Jones case we entirely avoid cutoff corrections by including all $N(N-1)/2$ pairwise interactions.
The finite-ranged polynomial potentials, with order-$N$ interactions, are much faster to analyze.
In our exploratory molecular dynamics simulations we enclosed a few hundred particles in a box
with quartic-potential very smooth boundaries and included a weak gravitational field. Our plan was to
observe phase boundaries directly.

\pagebreak

The polynomial potentials require simulation times of order $N$ rather than
$N^2$ for the force calculations. {\bf Figures 11 and 12} show six of the specimen potentials along with
eight of their static-lattice ``cold curves'', calculated for perfect triangular lattices. For the range of
densities shown all the shorter-ranged potentials, $\{ \ \phi^S_m \ \}$, have their minima at $r = 1$
with cold curve minima at a density $\sqrt{(4/3)} = 1.1547$ and a binding energy $e(\rho = 1.1547) = -3$.
Because the longer-ranged potentials extend to a separation of 2 their lattices are slightly compressed
from the ``close-packed'' density 1.1547. The binding energy is accordingly increased. See {\bf Figure 12}.

\begin{figure}
\includegraphics[width=1.5 in,angle = -90]{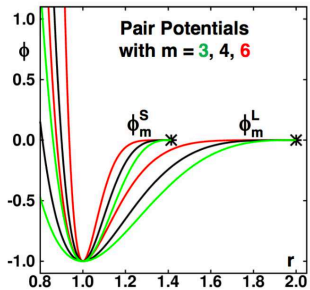}
\caption{
Two families of pair potentials, $\phi^L_m = (2-r)^{2m} - 2(2-r)^m$ and $\phi^S_m = (2-r^2)^{2m} - 2(2-r^2)^m$.
In the stress-free triangular lattice the short-ranged potentials, $\{ \ \phi^S_m(r<\sqrt{2}) \ \}$, have a
range $\sqrt{2}$ so that each particle only interacts with 6 nearest neighbors. In the longer-ranged case, with
$\phi^L_m(r<2)$, each particle interacts with 3 shells of 6 neighbors each.
}
\end{figure}

\begin{figure}
\includegraphics[width=1.5 in,angle = -90]{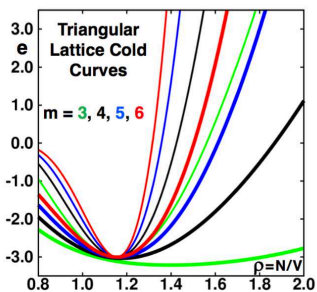}
\caption{
Cold curves for the triangular lattice with eight potentials of the two types shown in {\bf Figure 11}. The
stress-free square lattices are mechanically unstable to shear. The energies with wider bowls are drawn with
the wider lines. Those bowls, extending to $r=2$, include second and third neighbors at zero stress so that
the coresponding densities exceed the narrower-bowl value $\sqrt{(4/3)} = 1.1547$.
}
\end{figure}

\section{Gas Liquid Coexistence and the Mayers' ``Derby Hat''}

\begin{figure}
\includegraphics[width=1.5 in,angle = -90]{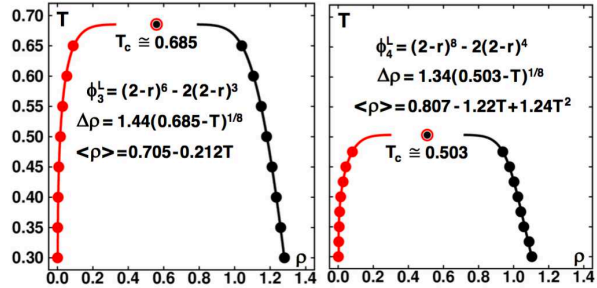}
\caption{
Coexisting densities for $\phi^L_3$ and $\phi^L_4$. Both cases closely reproduce Onsager's two-dimensional
Ising model power law dependence of the density difference, $\rho_{Liquid} - \rho_{Gas}$ on $T_c - T$. The
Mayers' 1940 ``Derby Hat'' idea for the region above the meniscus' disappearence appears in {\bf Figure 14}.
Each of the data shown here was generated with a million timesteps using $dt = 0.001$.
}
\end{figure}

{\bf Figure 13} shows our estimates for the gas-liquid coexistence curves for two of the short-ranged potentials.
These coexistence curves were obtained using an improved version of the ``liquid-ribbon'' method described by
Farid Abraham in 1980, using 256 particles. Here we use conventional molecular dynamics $(N = 3600, \ dt = 0.001,
\ t = 2000)$, with a cell, elongated horizontally and with periodic boundaries at its top and bottom.
Smooth particle averaging with Leon Lucy's weight function then provides density and pressure profiles. The
initial square-lattice configuration is quenched into the mechanically unstable region of the phase diagram
using Nos\'e-Hoover dynamics. The system rapidly separates into vapour and liquid phases. The density profile
is symmetrized prior to recording the values of the flat regions of the gas and liquid phases' densities. The
final configuration of each simulation becomes the starting point for the next-lower temperature simulation.
The full set of eight coexisting density pairs is then correlated with their temperatures, as shown in
{\bf Figure 13}.

The coexistence data are then fitted to the scaling laws shown in the figure. For the variation of the
density with temperature, $\Delta\rho(T)$, the exponents 1/7.5 and 1/7.9 resulted for the 6/3 and 8/4
long-ranged $(r \leq 2)$ potentials, $\phi^L_3$ and $\phi^L_4$. Because these exponents approximate
Onsager's 1/8, found analytically for the two-dimensional Ising model, we repeated the scaling law fits
using 1/8 for the exponent.

The ``Law of Rectilinear Diameters'' (the mean of gas and liquid densities varies linearly with temperature)
 was adequate for the 6/3 potential while a quadratic fit was needed in the righthand plot of the 8/4 data.
These choices enabled the vapour (red in {\bf Figure 13}) and liquid (black in that figure) densities to be
drawn in as continuous curves with critical temperatures of 0.685 and 0.503. We abandoned an effort to
estimate the critical region for the 10/5 potential, $\phi^L_5$. The relatively weaker binding energy indicated that weeks of
computer time might be required for an accurate assessment of that potential's critical point.

\begin{figure}
\includegraphics[width=1.5 in,angle = -90]{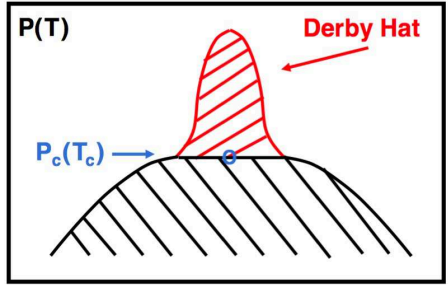}
\caption{
The Mayers' 1940 idea, now obsolete, that a meniscus-free region of infinite compressibility (shaped like
a ``Derby Hat'') would be found atop the two-phase coexistence region, is illustrated here in red. Our
simulations carried out above the critical temperature and with a small gravitational field  show the
presence of relatively large clusters of particles in the low-density upper portions of our containers,
as shown in {\bf Figure 15}.
}
\end{figure}

In 1940, as described in Chapter 14 of their {\it Statistical Mechanics} text\cite{b20}, Joseph and Maria Mayer
argued from the standpoint of their statistical-mechanical cluster theory that there is a highly-complex
``Derby-Hat'' critical region atop the coexistence curve, as is shown in {\bf Figure 14}. The red ``Hat''
region was thought to sit atop the two-phase region in which gas and liquid are separated by a meniscus. If
this construction were correct, as has been recently championed by Woodcock and Khmelinskii\cite{b21}, the
meniscus should suddenly vanish at the same temperature but at a whole range of different densities, different
by as much as ten percent according to the Mayers' estimate in three space dimensions. Here the main difference
seen above and below the apparent critical temperature of 0.685 for the $\phi^L_3$ potential is the
concentration of large clusters in the vapour phase, visible in {\bf Figure 15}.

These polynomial potentials avoid cutoff corrections. We briefly considered the usefulness of energy
comparisons with the square lattice, but preliminary calculations revealed that lattice unstable to
shears parallel to either the $x$ or the $y$ direction.  By contrast, as is consistent with hexagonal
symmetry, the triangular lattice has an isotropic shear modulus, a convenient property for modeling
atomistic results with isotropic continuum mechanics.

Either type of potential choice, $(2-r)^{12} - 2(2-r)^6$ or $(2-r^2)^6 - 2(2-r^2)^3$, can provide the
same curvature at the potential minimum of unity $\phi^{\prime \prime} = 72$ as does Lennard-Jones'
potential. {\bf Figure 16} shows pressure-density isotherms for all the potentials considered here. The
short-ranged potentials' isotherms have a positive compressibility (positive slope in the figure) while
Lennard-Jones' and the longer-ranged potentials have negative regions corresponding to the presence of
van der Waals' loops and gas-liquid coexistence.

\begin{figure}
\includegraphics[width=3. in,angle = -90]{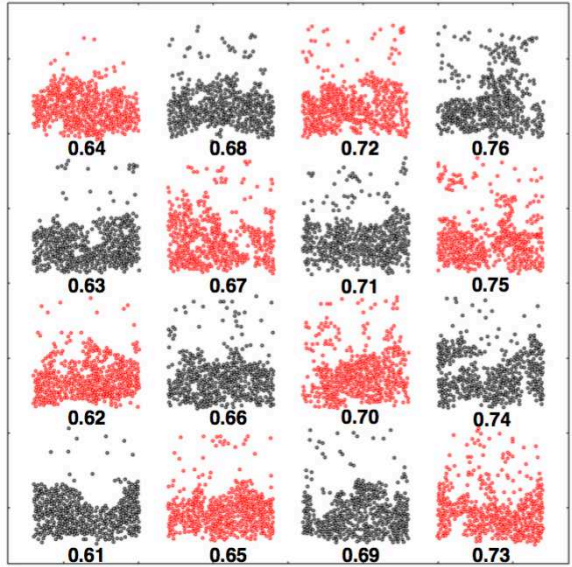}
\caption{
Sixteen snapshots of 400 particles with $\phi^L_3(r<2)$ controlled by a gravitational field $g = 0.01$.
The periodic width of the system is $\sqrt{800}$. Quartic repulsive potentials, $(1/4)dy^4$, apply beyond
the horizontal bounds $ | \ y \ | = \sqrt{200}$. The spacing between the snapshots is 20,000 timesteps,
corresponding to an elapsed time of 100. The lefthand columns, starting at lower left, correspond to
subcritical temperatures of 0.61 to 0.68 and the righthand columns to supercritical temperatures from 0.69
to 0.76, finishing at the upper right. The laptop time for all these simulations is a few minutes.
}
\end{figure}

\begin{figure}
\includegraphics[width=1.4 in,angle = -90]{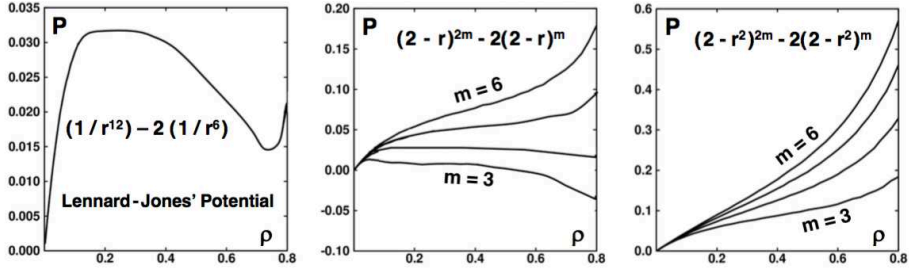}
\caption{
$T = 0.5$ isotherms for Lennard-Jones' potential and two families of polynomial potentials. The longer-ranged
potentials, $\{ \ \phi^L_m(r<2) = (2-r)^{2m} - 2(2-r)^m \ \}$, with m = 3, 4, 5, and 6, appear to exhibit
liquid phases while the nearest-neighbor shorter-ranged potentials,
$\{ \ \phi^L_m(r^2<2) = (2-r^2)^{2m} - 2(2-r^2)^m \ \}$, appear to go directly from the solid to the gas phase
on heating, with no intermediate liquid phase. Approximately 500 simulations were carried out for these
isotherms, all with $dt = 0.001$, one million timesteps, and Nos\'e-Hoover control of the temperature using
$\dot \zeta = 10[ \ (K_t/K_0) - 1 \ ]$. These data have been smoothed slightly for clarity.
}
\end{figure}

\section{Time-Averaged Thermostated Molecular Dynamics}

{\bf Figure 15} showed snapshots, equally-spaced in time and in temperature, of 400 longer-ranged
$\phi^L_3$ particles in a gravitational field.  The temperature increased stairstep-fashion from 0.61 in the
lower left corner to 0.76 at the upper right.  There are eight snapshots below the critical temperature and
eight above. Although the fluctuations are large it is clear that the cooler subcritical configurations are
qualitatively closer to a gas-liquid interface than the more diffuse supercritical configurations to the right.
The fluctuations evident in the snapshots can largely be removed by time averaging. {\bf Figures 17 and 18}
show density and pressure profiles with one million timesteps for each temperature. The eight subcritical
isotherms show a relatively sharp transition to a lower density gas-phase plateau.  Above the critical
temperature the density follows the barometer formula with a decreasing density and pressure with altitude.
The gravitational field organizes the fluid without noticeably perturbing the structure of the meniscus.

\begin{figure}
\includegraphics[width=2.8 in,angle = -90]{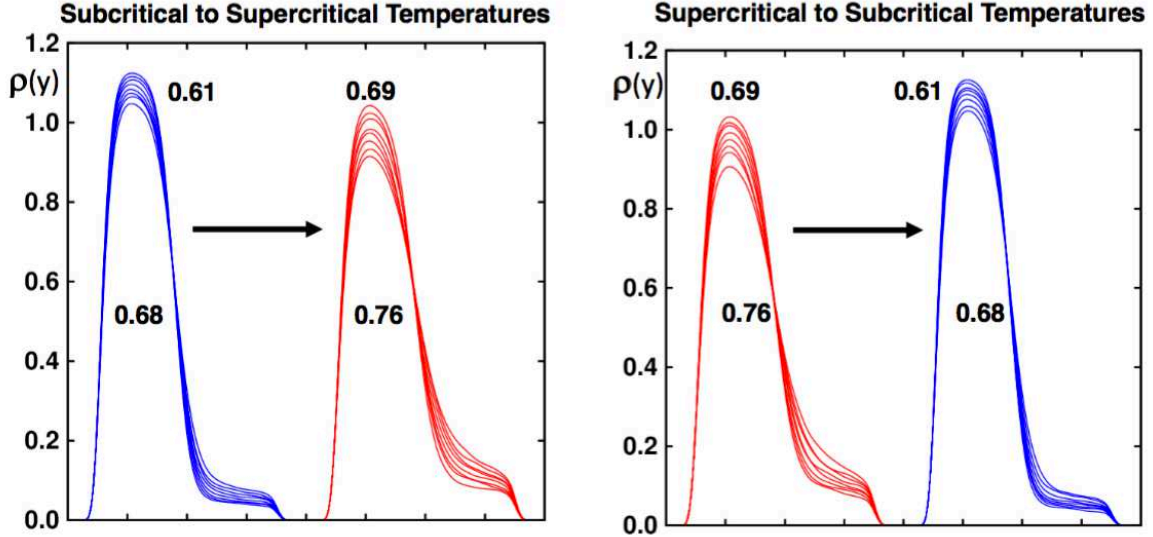}
\caption{
16 density profiles, subcritical in blue and supercritical in red, for 16 temperatures ranging from 0.61
to 0.76. The relatively sharp meniscus broadens noticeably at the critical temperature. Each profile is
the last-half average of two million Runge-Kutta timesteps with $dt = 0.005$. $N=400$ in a square
container, $L=\sqrt{800}$, periodic at the sides, quartic at the bottom and top, with a gravitational
field strength of 0.01.  Temperature increases at the left and decreases at the right.
}
\end{figure} 

\begin{figure}
\includegraphics[width=2.8 in,angle = -90]{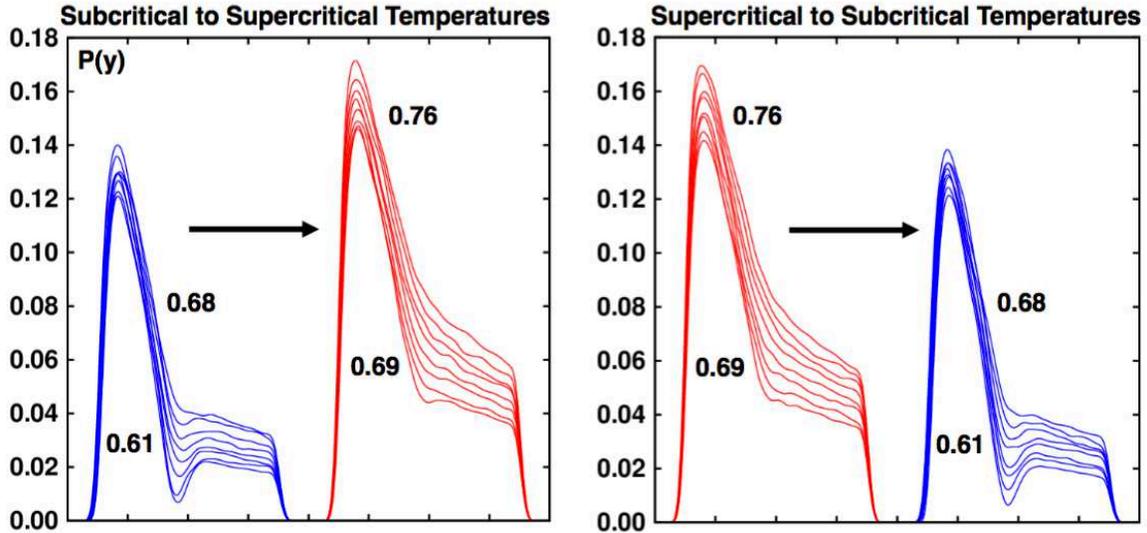}
\caption{
16 pressure profiles, subcritical in blue and supercritical in red, for 16 temperatures ranging from 0.61
to 0.76. The dip in the subcritical plots corresponds to the surface tension's negative contribution to the
mean pressure, $P = (P_{xx}+P_{yy})/2$. These data, just as those in {\bf Figure 17} required a bit under one
day of laptop time. $N=400$ in a square container, $L=\sqrt{800}$, periodic at the sides, quartic at the
bottom and top, with a gravitational field strength of 0.01. Temperature increases at the left and decreases
at the right.
}
\end{figure}

\section{The virial series, diffusion, and a movie}
In working to understand ``What is Liquid'' we found that gravity and time averaging were useful
computational tools. Gravity makes it possible to see an entire isotherm by stabilizing a
stationary equilibrium pressure gradient, $(\partial P/\partial y)_T = -\rho(y) g$.  Time
averaging makes it possible to reduce, and nearly to eliminate, the density and pressure
fluctuations which would otherwise obscure the structure of the liquid-gas meniscus. Here we
consider briefly the utility of three additional tools, the Mayers' virial series, local values
of the short-time anisotropic diffusion, and movies of the meniscus' evolution with relatively strong
fields.

\subsection{The Mayers' Virial Series}

In principle the Mayers' recipe for the virial coefficients $\{ \ B_n(T) \ \}$ provides an exact
route to the equation of state by formulating the pressure as a series in the density,
$$
PV/NkT = 1 + B_2\rho + B_3\rho^2 + B_4\rho^3 + B_5\rho^4 + ... \ .
$$
The efficient calculation of the higher coefficients in the Mayers' series has been greatly improved
by Richard Wheatley's work\cite{b22}. For hard spheres the fluid equation of state is well known from
computer experiments. Wheatley's 12-term series agrees with these experiments within a fraction of a
percent, all the way to freezing at (2/3) the close-packed density. Given that success one might well
expect that analogous calculations for our polynomial potentials would be useful near the critical
point. On the other hand, for $\phi^L_3$ at the critical point, the unit-distance maximum of the n-body
Mayer-function integrands,
$$
\prod f_{ij}(r_{ij}=1) \equiv \prod[e^{1/0.685} - 1] \simeq 3.3^{n^2/3} \ ,
$$
for large $n$, suggests convergence difficulties for the series. Numerical investigation of the first
five terms for $\phi^L_3(r<2)$ confirms this problem. Monte Carlo evaluation of the Mayer integrals with
$10^{11}$ configurations each gives the following results at the critical temperature, $T_c = 0.685$ :
$$
\{ \ B_2 = -4.25589 \ ; \ B_3 = -0.17_6 \ ; \ B_4 = 36._7 \ ; \ B_5 = 5_8 \ \} \ .
$$
Qualitatively the numbers are even less promising for $\phi^L_4(r<2)$ and for Lennard-Jones' 12/6
potential. Evidently the virial series will not help our understanding.  Even with the help of the
diminishing powers of the critical density, $0.560^{n-1}$, it is clear that the series is poorly
behaved. For $\phi^L_3$ at the critical point we find $PV/NkT = 1  - 2.38 - 0.06 + 6.45 + 6 \ ... \ ,$
which looks quite hopeless.  We conclude that the virial series is not a useful tool in the vicinity
of the critical point.

\subsection{Short-Time Diffusion and the Gas-Liquid Transition}

As the gas and liquid phases, with their different densities, have correspondingly different diffusion
coefficients it is worth investigating the anisotropicity of the diffusion induced by the gravitational
field's pressure gradient. Away from the meniscus we expect that the averaged mean-squared displacement
will be typical of the field-free displacements,
$$
\langle \ [ \ x(t)-x(0) \ ]^2 \ \rangle = \langle \ dx^2 \ \rangle \simeq
\langle \ [ \ y(t)-y(0) \ ]^2 \ \rangle = \langle \ dy^2 \ \rangle \ .
$$
In the dilute gas limit there is an apparent difference. At time $t$ and in the absence of collisions,
the mean-squared $x$ displacement is given by the Maxwell-Boltzmann's mean squared value, $(kT/m)t^2$
while the mean-squared $y$ coordinate, $\langle [ \ \dot y(0)t + (g/2)t^2 \ ]^2 \rangle = (kT/m)t^2 +
(g^2/4)t^4$ is bigger. In the dense solid, on the other hand, both displacements are tiny.

A more formal demonstration of the equilibrium distribution function in a field follows from the
Boltzmann equation for the probability density $f(t,y,p)$:
$$
(\partial f/\partial t) + p_y(\partial f/\partial y) -g(\partial f/\partial p_y) = 
{\rm collision \ term} \ .
$$
At low density the two-body collision term can be ignored. In the stationary state $(\partial
f/\partial t)$ vanishes. The other two terms cancel when $f$ includes $e^{-p^2/mkT}e^{-mgy/kT}$. All
that remains is the solution of a particle in the external field, $y(t) - y(0) = p_y(0)t - (1/2)gt^2$.
 
\begin{figure}
\includegraphics[width=4.0 in,angle = -90]{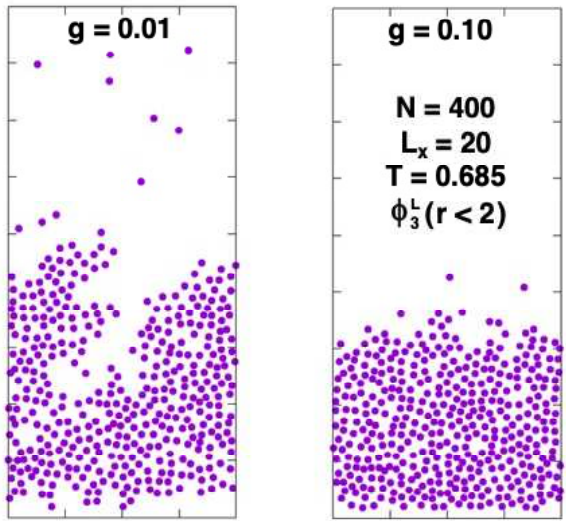}
\caption{
Typical isothermal equilibrium snapshots of 400 $\phi^L_3$ particles at the critical temperature
0.685 with gravitational field strengths of 0.01 (at the left) and 0.10 (at the right). Periodic
boundaries fix the system width at 20. A quartic boundary at the base supports the weight of the
system.
}
\end{figure}

\begin{figure}
\includegraphics[width=3.0 in,angle = -90]{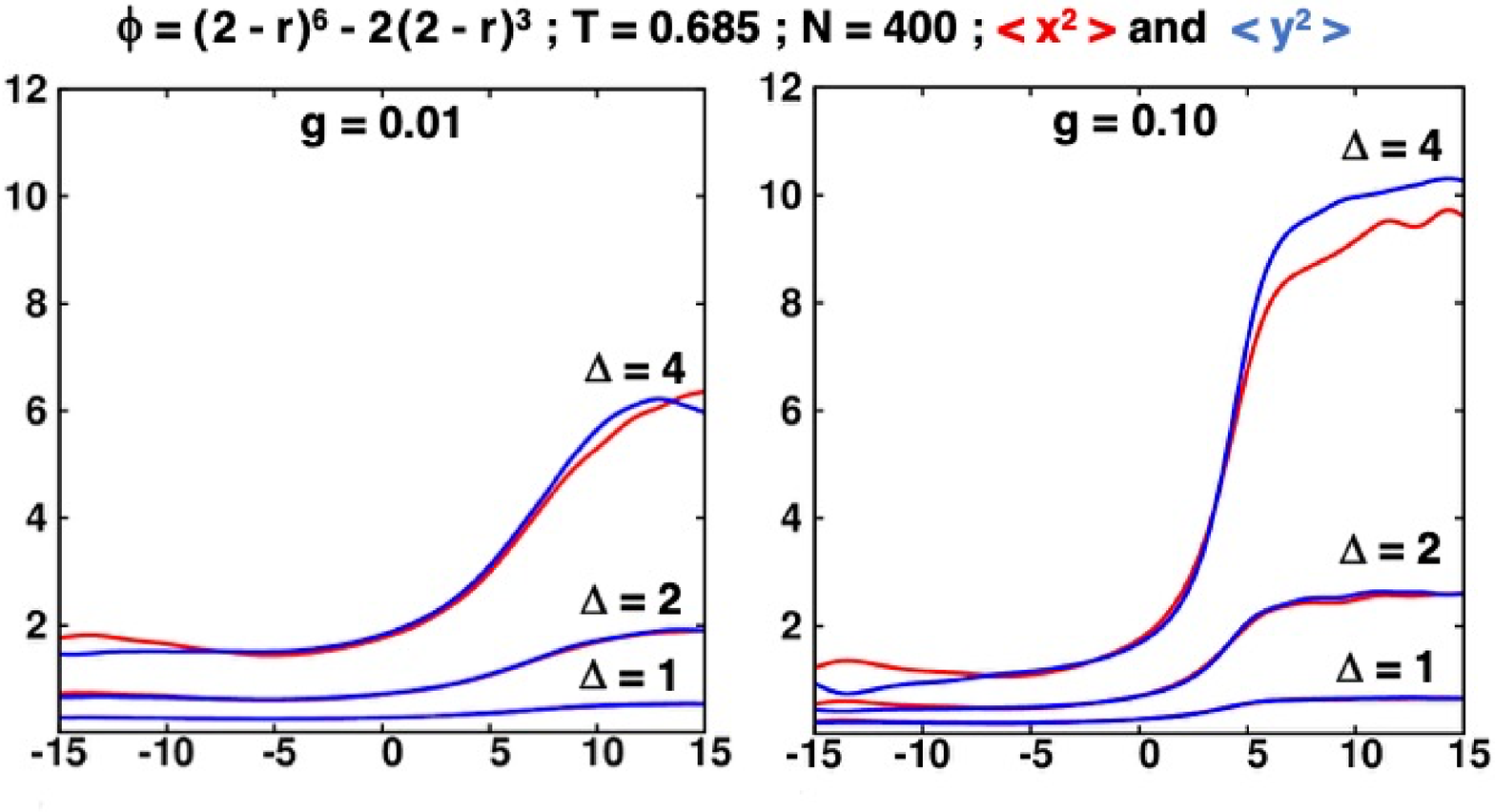}
\caption{
The mean-squared distances in the horizontal and vertical directions, $\langle \ x^2 \ \rangle$
and $\langle \ y^2 \ \rangle$ traveled in times $\Delta$ of 1, 2, and 4 are plotted as functions
of the midpoint $y$ coordinate. The boundary conditions are quartic, in a square box of sidelength
$\sqrt{800}$. At the left, corresponding to the weak field $ g = 0.01$ there is
no significant difference between the mean-squared horizontal and vertical displacements.
At the right there {\it is} a significant difference, with the vertical displacements, shown in
blue, larger, as
is consistent with collisionless kinetic theory. At the bottom of the container the easier $x$
motion can be seen clearly at the longest time, $\Delta = 4$.
}
\end{figure}

Consider two equilibrium systems under the influence of gravitational fields of 0.01 and 0.10,
illustrated by the late-time snapshots of {\bf Figure 19}. Notice particularly the difference in
the vapour pressures. For two similar systems, but with quartic boundaries, {\bf Figure 20} shows
the dependence of the diffusive mean-squared displacement on the sampling time $\Delta$. These
systems are held at their critical temperature, $T = 0.685$ subject to the same two field strengths,
$g =$ 0.01 and 0.10. The smooth-particle averages of $dx^2$ and $dy^2$ with vertical quartic
boundaries are based on 300 grid points with a smoothing length $h=2$. The $y$ coordinate associated
with these displacement pairs is the mean value, $(1/2)[ \ y(-\Delta t/2) + y(+\Delta t/2)$. In
{\bf Figure 20} the analysis of particle displacements as a function of the vertical coordinate is
carried out with a smoothing length $h= 2$ in a square system, $L = \sqrt{800}$. The corresponding
one-dimensional Lucy-function is
$$
w(z) = (5/4hL_x)(1 - 6z^2 + 8z^3 - 3z^4) \ ; \ z = | \ y_i - y_g \ |/2 \ {\rm for} \ z < 1 \ .
$$
Lucy's function
is used to calculate the ratios $\sum_i w_{ig} dx^2/\sum_i w_{ig}$ and $\sum_i w_{ig} dy^2/\sum_i w_{ig}$.
Here $w_{ig} \equiv w(| \ y_i - y_{grid} \ |)$. It is particularly interesting that the short-time
vertical displacement at the larger field is significantly greater than the horizontal, as predicted
by simple kinetic theory. An exploration of the details would make a rewarding research project. The state
dependence of the short-time diffusion is evidently a useful diagnostic tool.

\subsection{Movies of Isothermal Dynamics in a Gravitational Field}

In the early days of molecular dynamics computer-generated movies were staples at scientific 
meetings\cite{b23}. They showed not only the familiar motion of a gas filling a container, but also the
details of shear and heat flows.  These movies aided the intuition required to understand atomistic
dynamics from the standpoint of continuum mechanics. We expect now that such movies will help lead to
understanding the inhomogeneous correlations due to the gradients seen in drops, shockwaves, and
gravitational flows. Movies of two-dimensional systems with hundreds or thousands of particles
provide valuable insights while requiring very little computer time.  A sample movie prepared for
the readers of the arXiv and ``Computational Methods in Science and Technology'', can be found at
the website http://cmst.eu/wp-content/uploads/2021/03/sim-movie.mp4. {\bf Figure 21}, taken from a similar
 movie, shows at the left,
the early ``spinodal'' stage of a 400-particle $\phi^L_3$ system in a quartic box with $L=\sqrt{800}$.
The gravitational field, in {\bf Figure 21} a relatively large 0.2, supports only a tiny vapour pressure on
reaching the equilibrium state shown at the right.

\begin{figure}
\includegraphics[width=3.0 in,angle = -90]{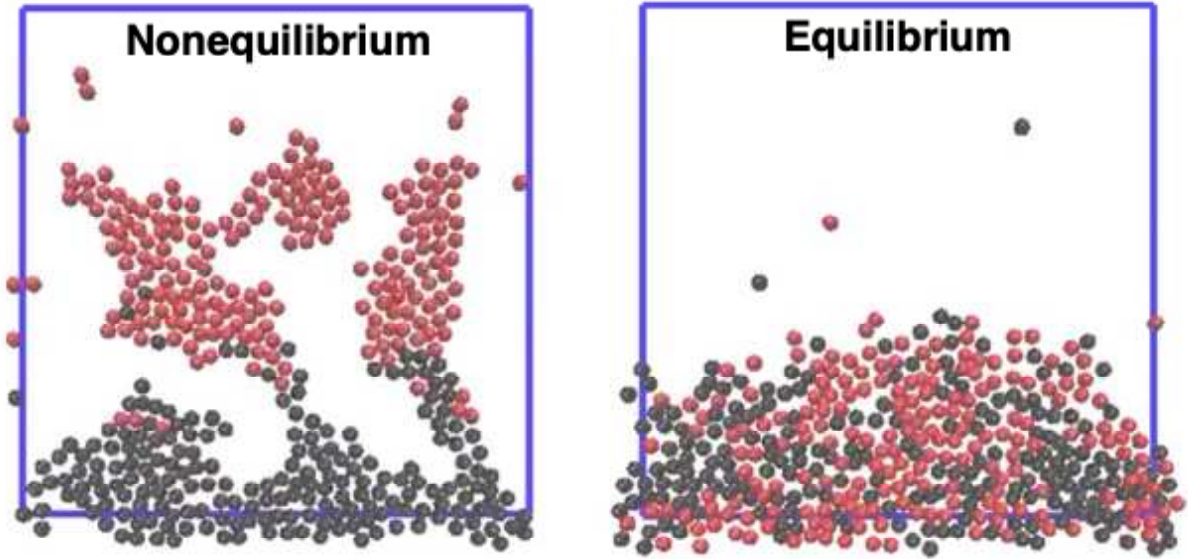}
\caption{
Two snapshots taken from a computer-generated movie illustrating ``spinodal decomposition'', at the
left, followed by thermal equilibration under the influence of a ``strong'' gravitational field,
$g = 0.20$. Initially the 400 particles in a square box of area 800 were arranged in an (unstable)
square lattice with the upper 200 colored red and the lower 200 black.  The pairwise-additive
interactions come from the $\phi^L_3$ potential.
}
\end{figure}

\section{Concluding Remarks}

We have explored the critical region for two varieties of polynomial pair potentials, finding that the
shorter-ranged family can sublime directly from the solid phase to the gas, without the intervention of
a liquid. The monotone nature of the isotherms of the shorter-ranged potentials, displayed in {\bf
Figure 15}  is consistent with this finding. The longer-ranged family, with second and third neighbor
interactions, form a liquid phase with a well-defined but fluctuating meniscus.  Time-averaging the
fluctuating profiles provides stable smooth estimates of the meniscus region separating the liquid and
gas.

With the longer-ranged potential $(2-r)^6 - 2(2-r)^3$ a visual inspection of the heated liquid in the
presence of a weak field reveals a complexity outside the normal range of thermodynamics. Clusters of
particles abound, from dimers and trimers up
to percolating clusters which stretch all the way across the simulation. In view of these fluctuating
features time-averaging is necessarily required to visualize and stabilize a liquid-gas interface. And
time averaging is not enough. In principle field-free time averaging would only produce a constant mean density
everywhere!

In order to ``see'' the definite boundary between liquid and gas we have considered an innovative version of
molecular dynamics, with a containerized region and a localizing gravitational field. This combination,
when time-averaged, provides density and pressure profiles in which the phases are separated by a meniscus. For small
field strengths these profiles resemble the Maxwell construction tie-line linking the two fluid phases
below the critical point.

This same technique is equally applicable using Monte Carlo simulations in the
canonical ensemble.  The presence of gravity provides a definite time-averaged interface, providing a
distinction  between the gas and liquid and addressing the ``What is Liquid'' directly, through Hannay's
interface. The complexities due to fluctuations moderated by surface tension can be overcome with gravity.
It is rewarding to see the subcritical dip in the pressure ({\bf Figure 18}) disappear at the estimated
critical temperature as calculated independently with an improved version of Abraham's ``liquid-ribbon'' technique. It is particularly
interesting to see that attractions beyond the first neighbors are needed for the liquid phase.

J\"urgen Schnack, in commenting on an earlier draft of this work, pointed out his analogous work (but with
a harmonic container) on the molecular dynamics of nuclear matter.  In particular that study\cite{b24}
estimated the critical temperature for hot oxygen at $10^{11}$ kelvins. We also thank John Ramshaw for his
cogent remarks on the applicability of hard-sphere perturbation theory to liquid-gas equilibria. 
 
This work achieved two goals.  First, we have introduced a useful simulation technique providing a
time-averaged description of the liquid-gas meniscus, curved or flat, depending upon the boundary
conditions, separating the two phases.  Second, we have introduced two families of pair potentials
which are relatively short-ranged, thereby avoiding the complications associated with cutoffs. These
models have shown that liquids are stabilized by interactions beyond nearest neighbors.  Our
simulations with gravity generate density gradients spanning a wide range of densities, including
the critical density.  The method developed here is relatively insensitive to the number of particles
chosen and the overall volume of the simulation container. No doubt it will suggest elaborations in
conjunction with other statationary processes, both at, and away from, equilibrium.

\pagebreak

\end{document}